\documentclass[11pt]{article}
\usepackage[dvips]{graphicx}

\begin{document}

\begin{center}
\bf{INVESTIGATION OF FLUORESCENCE RADIATION FOLLOWING RADIATIVE
RECOMBINATION OF IONS AND ELECTRONS} 
\end{center}

\vspace{5mm}
\begin{center}
A. Kupliauskien\.{e}\footnote{Corresponding author.
Tel.: +37052612723.\\E-mail address: akupl@itpa.lt
(A.~Kupliauskien\.{e}).}
\end{center}

\begin{center}
{\small\em Vilnius University Institute of Theoretical Physics 
       and Astronomy
A.Go\v{s}tauto 12, LT-01108 Vilnius, Lithuania}
\end{center}

\vspace{10mm}
{\bf Abstract}

A general expression for the double
differential cross section of fluorescence radiation
following photorecombination (PRF)
of polarized electrons and polarized
ions is derived by using usual atomic theory methods and is 
represented in the form of multiple expansions over spherical
tensors.
The ways of the application of the general expression suitable 
for the specific
experimental conditions are outlined by deriving the
asymmetry parameter  of angular distribution of 
PRF radiation in the case of nonpolarized ions and electrons.
This parameter is calculated for neon-like ions and bare nuclei.
A very strong dependence of the asymmetry parameter of
PRF radiation angular distribution 
on free electron energy is obtained.


\vspace{10mm}
{\bf 1. Introduction}

Radiative recombination (RR) is one of the dominant processes
leading to the decrease of ionization degree of highly charged ions
in laboratory and astrophysical plasmas.
In tokamak \cite{Fujimoto} and ion storage rings \cite{Stohlker}
the self alignment of recombined ions arises due to the directed
movement of ions that determines a strong polarization and
asymmetry of the angular distribution of fluorescence radiation.
The capture of an electron by an ion can take place in two
ways, i.e. photorecombination (PR) and dielectronic
recombination (DR).
In PR, an ion $A$ with charge $n+$ captures a free electron, and the
excess of the energy is emitted as PR radiation (PRR):
\begin{equation}
e^-({\rm\bf p}m_s) + A^{n+}(\alpha_1J_1M_1) \to  
A^{(n-1)+}(\alpha_2J_2M_2)
   + h\nu(\hat{\epsilon}_1{\rm\bf k}_{01}).
\label{eq:rr1}
\end{equation}
Here {\bf p} is the momentum of the electron, $m_s$ is its spin 
projection onto a chosen direction, $\hat{\epsilon}_1$ 
and {\bf k}$_{01}$ 
denote the polarization and wave vector of PRR, respectively, 
$\alpha_iJ_iM_i$ show other quantum numbers, total angular momentum 
and its projection of the ion $i$.

If a free electron is captured into a singly excited bound state, the 
recombined ion can decay by emitting fluorescence radiation (PRF)
\begin{equation}
 A^{(n-1)+}(\alpha_2J_2M_2) \to  A^{(n-1)+}(\alpha_3J_3M_3) +
h\nu(\hat{\epsilon}_2{\rm\bf k}_{02}).
\label{eq:rr2}
\end{equation}

DR is possible when the ion has one or more bound electrons.
Then a free electron is captured into an ion with simultaneous
excitation of a bound electron.
The resulting doubly excited state of an ion can decay both 
radiatively and via autoionization.
The DR occurs when the radiation (DRR) is emitted.
The experimental investigation of PR and PRF peculiarities is possible
because of fluorescence radiation emitted in the region of
wavelengths much longer in comparison with those of DRR.

The aim of the present work is the derivation of a general
expression for the differential cross section of PRF following
PR of polarized ions and polarized electrons by using
atomic theory methods [3-6]
that are alternative to the density matrix formalism \cite{Balashov}.

\vspace{10mm}
{\bf 2. Derivation of general expression}

The  double
differential cross section for PRF (\ref{eq:rr2}) following PR 
(\ref{eq:rr1}) process in two-step approximation can be written
in the form of multipole expansion over non-observed  states of 
the intermediate ion  \cite{K2004}:
$$
\frac{d^2\sigma(\alpha_1J_1M_1{\rm\bf p}m_s \to 
\alpha_2J_2 \hat{\epsilon}_{q_1} {\rm\bf k}_{01} \to
\alpha_3J_3M_3\hat{\epsilon}_{q_2} {\rm\bf k}_{02})}
{d\Omega_1 d\Omega_2} =
$$ 
\begin{equation}
\sum_{K_2N_2}
\frac{d\sigma_{K_2N_2}(\alpha_1J_1M_1{\rm\bf p}m_s \to 
\alpha_2J_2 \hat{\epsilon}_{q_1} {\rm\bf k}_{01})}
{d\Omega_1}
\;
\frac{dW_{K_2N_2}(\alpha_2J_2 \to
\alpha_3J_3M_3\hat{\epsilon}_{q_2} {\rm\bf k}_{02})}
{d\Omega_2}.
\label{eq:rr3}
\end{equation}
Here $d\sigma/d\Omega$ represents PR differential cross section
that is inverse to that of photoionization.
Its expression can be written by using the Milne \cite{Sobelman}
relation and its multipole expansion is \cite{KT2003,K2004}:
$$
\frac{d\sigma_{K_2N_2}(\alpha_1J_1M_1{\rm\bf p}m_s \to 
\alpha_2J_2 \hat{\epsilon}_{q_1} {\rm\bf k}_{01})}
{d\Omega_1}
$$
$$
=
\sqrt{4\pi}\frac{\pi \alpha^2 E^2_1}{\varepsilon}
\sum_{K_1,K_r,K_\lambda,K_s,K_j,K,k_1,k_1'}
{\cal B}^{ph}(K_2,K_1,K_r,K_\lambda,K_s,K_j,K,k_1,k'_1)
$$
$$
\times
\left[2K_2+1\right]^{1/2}\sum_{N_1,N_r,N_\lambda,N_s,N_j,N}
\left[\begin{array}{ccc}
K_1&K_j&K\\N_1&N_j&N\end{array}\right]
\left[\begin{array}{ccc}
K_\lambda &K_s&K_j\\N_\lambda&N_s&N_j\end{array}\right]
\left[\begin{array}{ccc}
K_2&K_r&K\\N_2&N_r&N\end{array}\right]
Y^*_{K_\lambda N_\lambda}(\hat{p})
$$
\begin{equation}
\times
T^{*K_1}_{N_1}(J_1,J_1,M_1|\hat{J}_1)
T^{*K_s}_{N_s}(s,s,m_s|\hat{s})
T^{K_r}_{N_r}(k_1,k'_1,q_1|\hat{k}_{01}).
\label{eq:rr4}
\end{equation}
Here $\varepsilon$ is the energy of the free electron in atomic units, 
$E_1$ is the energy of PRR in atomic units, 
$\alpha$ is the fine structure constant,
$a_0$ is the Bohr radius, and 
\begin{equation}
T^K_N(J,J',M|\hat{J})
=(-1)^{J'-M}\left[\frac{4\pi}{2J+1}\right]^{1/2}
\left[\begin{array}{ccc}
J & J' & K\\M & -M & 0\end{array}\right]
Y_{KN}(\hat{J}).
\label{eq:rr4_1}
\end{equation}

The expression for the differential probability of PRF
is as follows \cite{KT2004}:
$$
\frac{dW_{K_2N_2}(\alpha_2J_2 \to
\alpha_3J_3M_3\hat{\epsilon}_{q_2} {\rm\bf k}_{02})}
{d\Omega_2}
=
\frac{\alpha^2E^2_2}{2\pi} \sum_{K'_r,K_3,k_2,k'_2}
\frac{\sqrt{2K_2+1}}{2J_2+1}
{\cal A}^r(K_2,K'_r,K_3,k_2,k'_2)
$$
\begin{equation}
\times
\sum_{N'_r,N_3}
\left[\begin{array}{ccc}
K_2&K'_r&K_3\\N_2&N'_r&N_3\end{array}\right]
T^{K_2}_{N_2}(J_3,J_3,M_3|\hat{J}_3)
T^{K'_r}_{N'_r}(k_2,k'_2,q_2|\hat{k}_{02}).
\label{eq:rr5}
\end{equation}
Here $E_2$ is the energy of the fluorescence photon in atomic units.

The expressions for 
${\cal B}^{ph}(K_2,K_1,K_r,K_\lambda,K_s,K_j,K,k_1,k'_1)$ \cite{KT2003}
and
${\cal A}^r(K_2,K'_r,K_3,k_2,k'_2)$ \cite{KT2004} 
are:
$$
{\cal B}^{ph}(K_2,K_1,K_r,K_\lambda,K_s,K_j,K,k_1,k'_1)=
 \sum_{\lambda,j,J,\lambda',j',J'}(2J+1)(2J'+1)
(-1)^{\lambda'}
$$
$$
\times
\langle \alpha_1 J_1\varepsilon_1\lambda(j)J||Q^{(k_1)}||\alpha_0
J_0\rangle
\langle \alpha_1 J_1\varepsilon_1\lambda'(j')J'||Q^{(k'_1)}||
\alpha_0J_0\rangle^*
$$
$$
\times
[(2J_1+1)(2K_j+1)((2J_2+1)(2k_1+1)(2s+1)(2\lambda+1)
(2\lambda'+1)(2j+1)(2j'+1)]^{1/2}
$$
\begin{equation}
\times
         \left[
\begin{array}{ccc}
\lambda &\lambda' & K_\lambda\\ 0 & 0 & 0
\end{array}
          \right]
         \left\{
\begin{array}{ccc}
J_1 &K_1 & J_1\\ k_1 & K_r & k'_1\\ J' & K & J
\end{array}
          \right\}
         \left\{
\begin{array}{ccc}
J_2 &K_2 & J_2\\ j' & K_j & j  \\ J' & K & J
\end{array}
          \right\}
         \left\{
\begin{array}{ccc}
\lambda' &K_\lambda & \lambda\\ s & K_s & s  \\ j' & K_j & j
\end{array}
          \right\},
\label{eq:rr6}
\end{equation}
$$
{\cal A}^r(K_1,K'_r,K_2,k_1,k'_2) =
( \alpha_2J_2 ||Q^{(k_2)}||\alpha_1J_1) 
( \alpha_2J_2 ||Q^{(k'_2)}||\alpha_1J_1)^*
$$
\begin{equation}
\times
\left[\frac{(2K_1+1)(2J_2+1)(2k_2+1)}{2K_2+1}\right]^{1/2}
\left\{\begin{array}{ccc}
J_1 & K_1 &J_1 \\ k_2 & K'_r & k'_2 \\ J_2 & K_2 & J_2
\end{array}\right\} .
\label{eq:rr7}
\end{equation}

The reduced matrix element in (\ref{eq:rr6}) and (\ref{eq:rr7})
has the expression \cite{KT2004}:
\begin{equation}
\langle \alpha_2 J_2||Q^{(k)}||\alpha_1 J_1\rangle
=  k_{0}^{k-1/2} \sum_{p=0,1} \left[\frac{k+1}{k}\right]^{1/2}
\frac{i^{k}}{(2k-1)!!}
\langle \alpha_2 J_2|{\cal Q}^p_{k}|\alpha_1 J_1\rangle.
\end{equation}
For the electrical multipole (E$k$) transitions, $p=0$, and
the transition operator in (9) is  \cite{KT2004}
\begin{equation}
{\cal Q}^0_{kq}= - r^k C^{(k)}_q
\end{equation}
and, for the magnetic multipole transition (M$k$)($p=1$), it is
\begin{equation}
{\cal Q}^1_{kq}
= \frac{ iq}{c} [k(2k-1)]^{1/2}r^{k-1}
\left\{ \frac{1}{k+1}\left[ C^{(k-1)} \times  L^{(1)}
\right]^{(k)}_q
+\left[C^{(k-1)} \times S^{(1)}\right]^{(k)}_q\right\}.
\end{equation}
\noindent
Here  $L$ and $S$ are the operators of the orbital and
spin  angular momentum, respectively,
$C^{(k)}_q$ is the operator of the spherical function normalized to
$[4\pi/(2k+1)]^{1/2}$.

In nonrelativistic approximation,
the expression (\ref{eq:rr3}) represents the most general case
of the double differential cross section for the PRF process of
polarized ions and polarized electrons.
In two-step approximation, it enables us to obtain information
about the polarization, asymmetry of the angular distributions
of PRR and PRF as well as angular correlations between PRR and
PRF.
The parameters describing the polarization of the intermediate
 and final ions can also be derived.

\vspace{10mm}
{\bf 3. Special cases}

The cases of PRF following PR of nonpolarized and polarized or
aligned ions with nonpolarized electrons are of importance
for tokamak plasmas.
To obtain the expression for differential cross section
describing the angular distribution of PRF radiation when
the PRR and final states of recombined ion and polarization
of PRF radiation are not registered, one needs to perform
the summation in  (\ref{eq:rr3})
over $M_3$, polarization of radiation,
$q_2=q_1=\pm 1$, and integration over the angles of PRR.
The $\sum_{M} T^K_N(J,J,M|\hat{J})=\delta(K,0)\delta(N,0)$.
The integration gives
$\int d\Omega Y_{KN}(\theta,\phi)=\delta(K,0)\delta(N,0)$.
The zero ranks of the corresponding multipole expansion
tensors should be inserted into (\ref{eq:rr4}) and (\ref{eq:rr5})
that leads to the simplification of (\ref{eq:rr3}).
For the randomly oriented both ions and electrons, 
the expression (\ref{eq:rr3}) should be averaged with respect to
the states of the ion and electron.

Thus, in the of PR of nonpolarized ions and electrons and the choice
of the laboratory $z$ axis along the direction of electrons,
the expression for the differential cross section of PRF can
be written as follows:
$$
\frac{d\sigma(\alpha_1J_1 \to \alpha_2J_2  \to
\alpha_3J_3 {\rm\bf k}_{02})} {d\Omega_2} =
\frac{1}{2(2J_1+1)}
\sum_{M_1,M_3,m_s,q_1,q_2}
$$
$$
\times
\int d\Omega_1
\frac{d^2\sigma(\alpha_1J_1M_1{\rm\bf p}m_s \to 
\alpha_2J_2 \hat{\epsilon}_{q_1} {\rm\bf k}_{01} \to
\alpha_3J_3M_3\hat{\epsilon}_{q_2} {\rm\bf k}_{02})}
{d\Omega_1 d\Omega_2}
$$ 
\begin{equation}
=\sigma(\alpha_1J_1\to \alpha_2J_2)
\frac{W(\alpha_2J_2 \to \alpha_3J_3)}{4\pi}
\left[1+\sum_{K_r>0} 
\beta_{K_r}P_{K_r}(\cos \theta)\right].
\label{eq:rr8}
\end{equation}
$K_r$ acquires even values. 
For electrical dipole radiation $k_2=k'_2=1$, $K_r=0,2$, and
\begin{equation}
\beta_2=\alpha A_2
\label{eq:rr9}
\end{equation}
where
\begin{equation}
\alpha=
\left[\frac{5(2J_2+1)}{2}\right]^{1/2}
\frac{{\cal A}^r(2,2,0,1,1)}{{\cal A}^r(0,0,0,1,1)}
= (-1)^{J_2+J_3+1}
\left[\frac{3(2J_2+1)}{2}\right]^{1/2}
\left\{\begin{array}{ccc}J_2&J_2&2\\1&1&J_3\end{array}\right\}
\label{eq:rr10}
\end{equation}
depends on only the quantum numbers of recombined ion.

The parameter $A_2$ is the alignment of recombined ion, and its
expression is:
\begin{equation}
A_2=\frac{5}{\sqrt{2J_2+1}}
\frac{{\cal B}^{ph}(2,0,0,2,0,2,2,1,1)}
{{\cal B}^{ph}(0,0,0,0,0,0,0,1,1)}.
\label{eq:rr11}
\end{equation}
A very strong dependence of $A_2$ on free electron energy can
be expected because of the sum over $\lambda,\lambda'$ in
${\cal B}(2,0,0,2,0,2,2,1,1)$, i.e. interference of terms
exists, that differs from the alignment $A_2$ of photoionization
where $\lambda=\lambda'$ indicating the absence of interference
terms.

In the case of PR of polarized
ions and nonpolarized electrons, the expression for the
differential cross section describing the angular distribution of
PRF radiation is as follows:
$$
\frac{d^2\sigma(\alpha_1J_1M_1 \to \alpha_2J_2  \to
\alpha_3J_3 {\rm\bf k}_{02})} {d\Omega_2} 
$$
$$
=
\frac{1}{2}
\sum_{M_3,m_s,q_1,q_2}
\int d\Omega_1
\frac{d^2\sigma(\alpha_1J_1M_1{\rm\bf p}m_s \to 
\alpha_2J_2 \hat{\epsilon}_{q_1} {\rm\bf k}_{01} \to
\alpha_3J_3M_3\hat{\epsilon}_{q_2} {\rm\bf k}_{02})}
{d\Omega_1 d\Omega_2}
$$ 
\begin{equation}
=(2J_1+1)\sigma(\alpha_1J_1\to \alpha_2J_2)
\frac{W(\alpha_2J_2 \to \alpha_3J_3)}{4\pi}
\left[1+\sum_{K_2>0,N_2} \beta_{K_2} Y_{K_2N_2}(\theta_2,\phi_2)
A_{K_2N_2}(\theta_A,\phi_A)\right].
\label{eq:rr12}
\end{equation}
Here the laboratory $z$ axis is aligned along the direction of free
electrons.
In (\ref{eq:rr12}), 
\begin{equation}
\beta_{K_2}=\sum_{k_2,k'_2}
(-1)^{k'_2-q_2}\left[\frac{4\pi}{2k_2+1}\right]^{1/2}
\left[\begin{array}{ccc}k_2&k'_2&K_2\\q_2&-q_2&0\end{array}\right]
\frac{{\cal A}^r(2,2,0,1,1)}{{\cal A}^r(0,0,0,1,1)},
\label{eq:rr13}
\end{equation}
and differential alignment is defined as
$$
A_{K_2N_2}(\theta_A,\phi_A)=\sum_{K_1,K_\lambda,k_1}
(-1)^{K_2-N_2+J_1-M_1}
\left[\frac{4\pi(2K_2+1)(2K_\lambda+1)}{2J_1+1}\right]^{1/2}
\left[\begin{array}{ccc}J_1&J_1&K_1\\M_1&-M_1&0\end{array}\right]
$$
\begin{equation}
\times
\frac{1}{2k_2+1}
\left[\begin{array}{ccc}K_1&K_\lambda&K_2\\N_2&0&N_2
\end{array}\right]
\frac{{\cal B}^{ph}(K_2,K_1,0,K_\lambda,0,K_\lambda,K_2,k_1,k_1)}
{{\cal B}^{ph}(0,0,0,0,0,0,0,k_1,k_1)}
Y^*_{K_1N_2}(\theta_A,\phi_A).
\label{eq:rr14}
\end{equation}
The angles in (\ref{eq:rr12}) and (\ref{eq:rr14}) are measured
from the direction of free electron.

In electrical dipole approximation, $k_1=k'=1$, $K_2=0,2$, and the
expression for $\beta_2$ is defined by (\ref{eq:rr9}).

\vspace{10mm}
{\bf 4. Calculations}

The asymmetry parameters $\beta_2$ (\ref{eq:rr9}) of the angular
distribution for PRF in the case of PR of bare nuclei and 
nonpolarized Ne-like ions Na$^+$, Mg$^{2+}$, Al$^{3+}$, Ar$^{8+}$,
Fe$^{16+}$, and Zn$^{20+}$ with nonpolarized electrons were
calculated.
The reduced matrix elements in ${\cal B}^{ph}(2,0,0,2,0,2,2,1,1)$
(\ref{eq:rr6}) were calculated by using a computer program for
the photoionization of atoms \cite{K2001}.
For the calculation of (\ref{eq:rr9}) -- (\ref{eq:rr11}),
a computer program was created in the present work.

In the calculations, the energies of a free electron are chosen
from 0 up to 450~eV that allows the use of dipole approximation
for PRR.
The energy of PRR does not exceed 1.5~keV for Al$^{18+}$ and 
Zn$^{20+}$, therefore, it can be expected that the 
contribution of higher multipoles does not exceed 2\% for PR
cross sections.

In the case of bare nuclei of He, F, and Ar atoms,
calculated asymmetry parameters $\beta_2$ of the angular distribution
of PRF for the transition 2p $^2$P$_{3/2} \to$ 1s~$^2$S$_{1/2}$ 
are shown in Fig.~1.
The relativistic calculations for Xe$^{54+}$ \cite{Eichler}
are also presented for comparison.
The values of $\beta_2$ are very similar near zero energies
of free electron and decrease down from 0.34 with increasing
electron energy.
For higher free electron energies, the values of $\beta_2$
increase if the nuclear charge increases.

For PRF of Ne-like ions via transition
2p$^6$ $^1$S$_0 \to$ 2p$^6$3p $^2$P$_{3/2} \to$ 2p$^6$3s $^2$S$_{1/2}$,
the values of calculated asymmetry parameters $\beta_2$
are presented in Fig.~2.
A very strong variation of $\beta_2$ on free electron energy can
be noticed for ions of low ionization degree, i.e. Na$^+$ and
Mg$^{2+}$ which gradually disappears with increasing nuclear charge.
The minimum  noticed for Na$^+$ and Mg$^{2+}$ coincides with the
Cooper minimum in photoionization of Na and Mg$^+$ that is
reversed to PR.
The strong dependence of $\beta_2$ on free electron energy is
defined by the alignment $A_2$ (\ref{eq:rr11}) of the recombined
ion.
The strong variations of $A_2$ on free electron energy
occurs due to the interference of $\lambda,\lambda'$ in
${\cal B}^{ph}(2,0,0,2,0,2,2,1,1)$.

{\bf Figure captions:}

Fig. 1. Asymmetry parameter $\beta_2$ of the angular distribution of
fluorescence   $2p ^2P_{3/2} \to 1s ^2S_{1/2}$ radiation
following photorecombination of bare nuclei vs free electron
energy. Calculations for  Xe$^{54+}$ are taken from \cite{Eichler}.

Fig. 2. Asymmetry parameter $\beta_2$ of the angular distribution of
fluorescence   $3p ^2P_{3/2} \to 1s ^2S_{1/2}$ radiation
following photorecombination of neon-like ions vs free electron
energy.

Figure~1

\begin{figure}[h!]
\begin{center}
\includegraphics*[width=14cm,height=10cm]{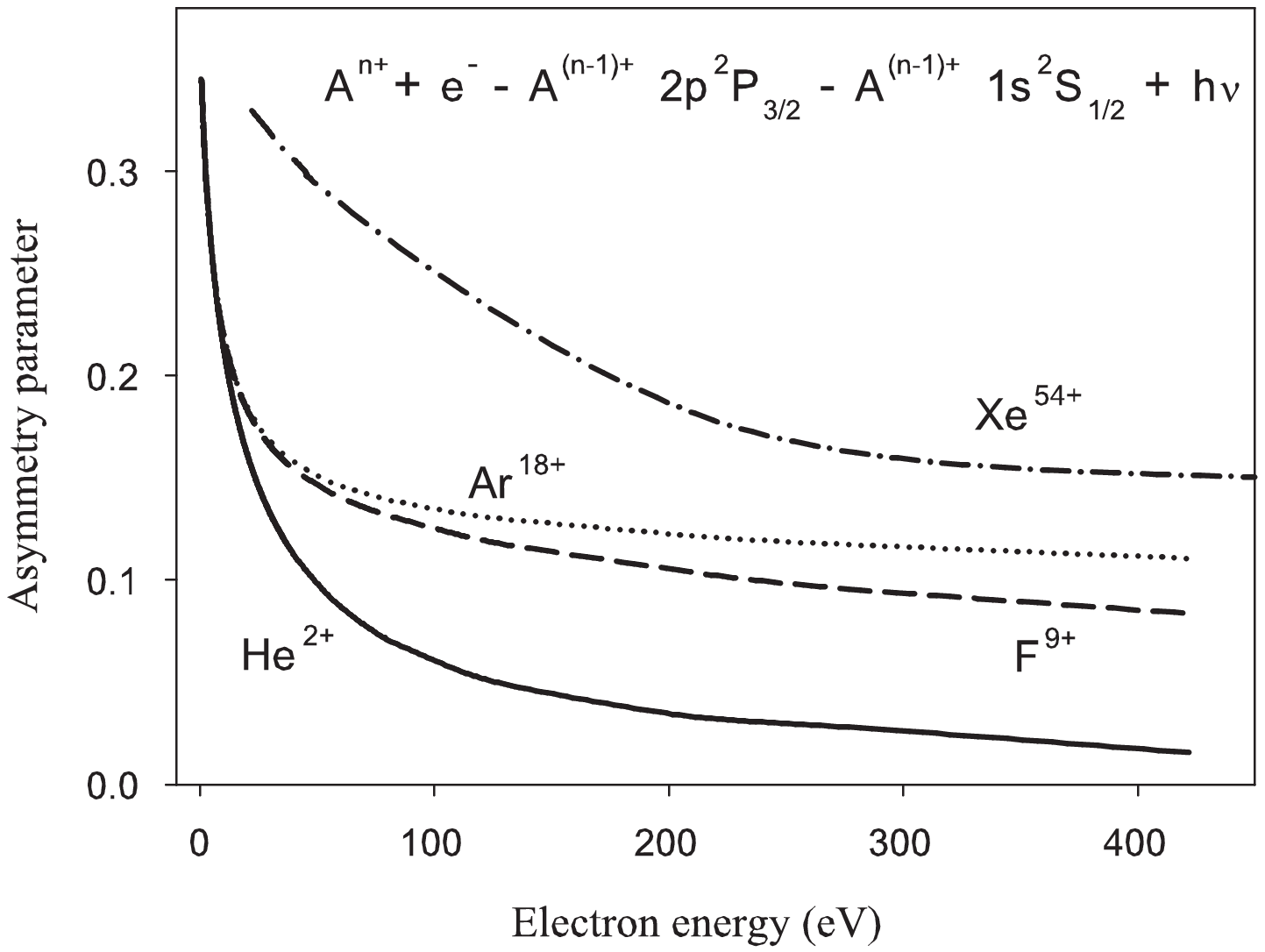}
\end{center}
\end{figure}

\clearpage
Figure~2

\begin{figure}[h!]
\begin{center}
\includegraphics*[width=14cm,height=10cm]{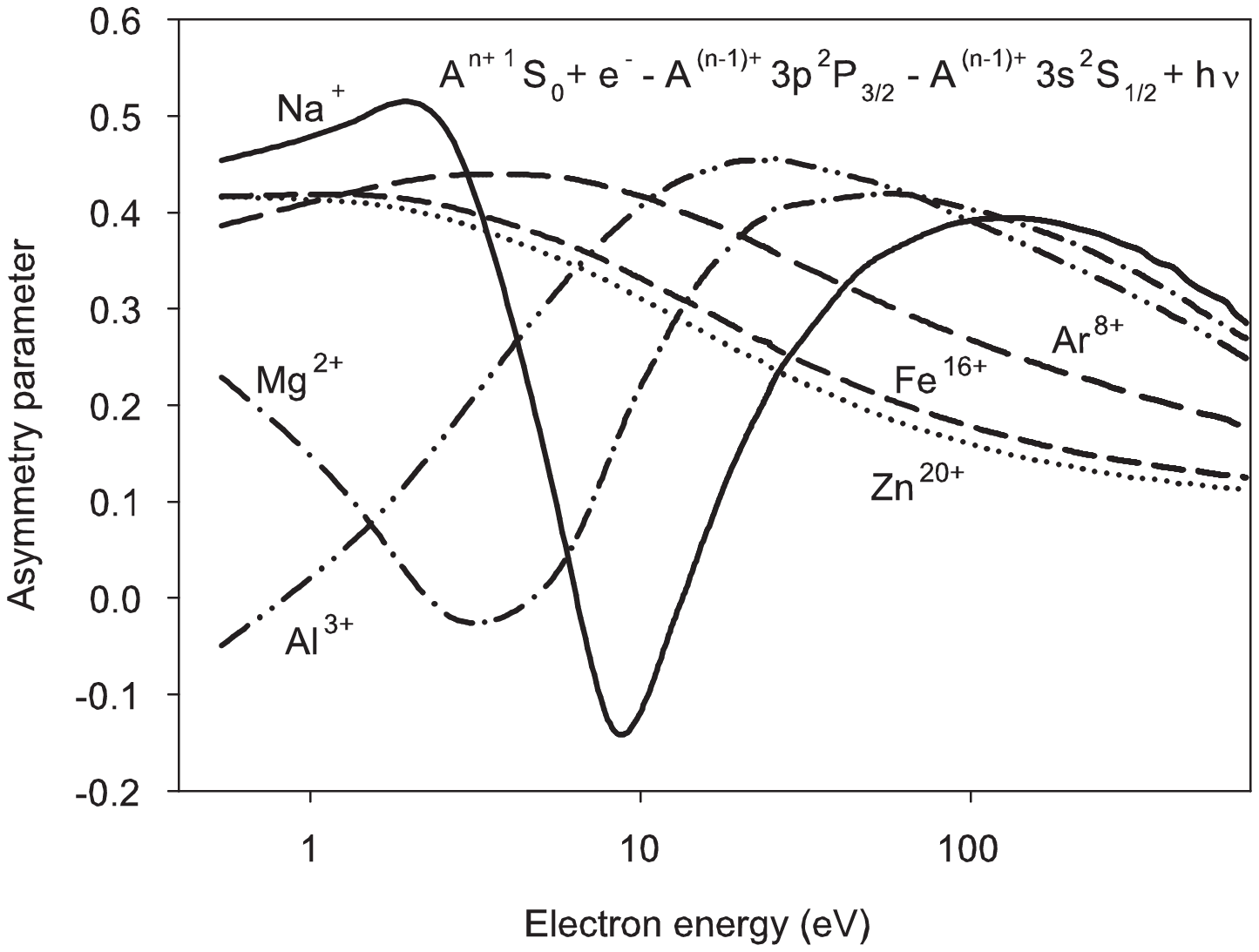}
\end{center}
\end{figure}

\end{document}